# Perpendicular-current Studies of Electron Transport Across Metal/Metal Interfaces.

W.P. Pratt Jr. and J. Bass

Department of Physics and Astronomy, Michigan State University, East Lansing, MI 48824 USA.

Abstract

We review what we have learned about the scattering of electrons by the interfaces between two different metals (M1/M2) in the current-perpendicular-to-plane (CPP) geometry. In this geometry, the intrinsic quantity is the specific resistance, AR, the product of the area through which the CPP current flows times the CPP resistance. We describe results for both non-magnetic/non-magnetic (N1/N2) and ferromagnetic/non-magnetic (F/N) pairs. We focus especially upon cases where M1/M2 are lattice matched (i.e., have the same crystal structure and the same lattice parameters to within ~ 1%), because in these cases no-free-parameter calculations of 2AR agree surprisingly well with measured values. But we also list and briefly discuss cases where M1/M2 are not lattice matched, either having different crystal structures, or lattice parameters that differ by several percent. The published calculations of 2AR in these latter cases do not agree so well with measured values.

1. Introduction and Overview.

In multilayered structures, composed of alternating layers of different metals (M1/M2), scattering of electrons at interfaces can have a major impact on the transport. In this review we concentrate upon the contribution of such scattering to the electrical resistance in the current-perpendicular-to-plane (CPP) geometry. In this geometry, the intrinsic quantity is the specific resistance, AR, the product of the area through which the CPP current flows times the CPP resistance. Since each bilayer of M1/M2 contributes two interfaces, we focus upon 2AR. For non-magnetic/non-magnetic (N1/N2) metal pairs, 2AR completely describes the interfacial scattering of present interest. For ferromagnetic/non-magnetic (F/N) metal pairs, two quantities are needed. The two most convenient for our purposes are the interfacial specific resistance, $2AR_{F/N}^*$, and the asymmetry parameter, $\gamma_{F/N}$, both defined below.

Three techniques have been used to measure the CPP-R of samples: (1) Short, wide multilayers sandwiched between superconducting Nb cross-strips [1]; (2) long, thin multilayer wires electrodeposited into tubular holes in plastic or $Al_2O_3$ [2-4], and (3) micro- or nano-pillared multilayers with widths and lengths of roughly comparable magnitudes, shaped by optical or electron-beam lithography [5]. We focus upon results found using technique (1). We chose this technique for three reasons: (a) it is our own work; (b) it has so far provided the most reliable values of 2AR, due to generally superior control over the area A and the contact resistance; (c) it has allowed the widest range of metal pairs to be studied. Moreover, it has produced all of the published, directly obtained values of $2AR_{N1/N2}$. Its limitation is its restriction to cryogenic temperatures (typically 4.2K). Fortunately, this temperature limitation is not a problem for comparing with calculations, which are valid in the 0K limit. What is known about the temperature dependence of 2AR comes from a combination of techniques (2) and (3). Published evidence suggests that 2AR is usually only weakly temperature dependent from 4.2K to 293K—e.g., changing by $\leq 20\%$ [6,7].

The review is organized as follows. This introduction provides a brief overview. Section 2 describes the experimental techniques used to obtain the results that are later presented. Section 3 specifies in detail the multilayer structures used to find $2AR_{N1/N2}$, and outlines the more complex procedures used to find $2AR_{F/N}^*$. This section also contains figures showing data for finding $2AR_{N1/N2}$. Section 4 reviews the main features of quantitative calculations of $2AR_{N1/N2}$ and $2AR_{F/N}^*$. Section 5 contains the collected results, mainly in two tables of measured and calculated values [7-24]. Table 1 is for metals M1/M2 that are lattice matched—i.e. that have the same crystal structures, and lattice parameters that agree to within ~ 1%. In this case, no-free-parameter calculations give surprisingly good agreement with the measured values. Table 2 is for metals that are not lattice matched—either their lattice parameters differ by well over 1%, or else they have different crystal structures (e.g., face-centered-cubic (fcc) versus body-centered-cubic (bcc)). Together, low and high angle x-ray studies [12], and cross-sectional transmission electron microscopy (TEM) studies [22], show that the sputtered multilayers are well layered, growing in columns with typical 20-50 nm caliper dimensions in-plane, with the layers in close-packed planes--(111) for fcc and (110) for bcc. The normals to these planes give the current directions assumed for the calculations described in section 4 below. For lattice matched pairs, the columns grow locally almost epitaxially. For non-lattice matched pairs, the interfaces adjust locally via dislocations and plane tilting [22]. The x-ray, TEM, and resistance results all indicate that the interfaces are intermixed, typically over 3-4 monolayers (ML), with some interfacial wavyness, especially for non-lattice matched pairs [22]. Section 6 is a summary and conclusions.

2. Experimental Techniques for Sample Preparation and Measuring.

Details of the sample preparation and measuring techniques are given in refs. [8] and [25]. The multilayers of interest, sandwiched between crossed Nb strips, are produced by sputtering in a chamber with 6 sputtering targets and an in-situ mask changing system that allows sequential rotation of a mask with one closed position and three differently shaped open ones, so that the whole sample can be sputtered without breaking vacuum. To maintain sample purity, the sputtering system is initially pumped down, baked to ~ 100°C, and then liquid nitrogen is flowed through an internal Meissner trap to freeze out any residual water vapor. The background pressure after baking and cooling is usually $\leq 2 \times 10^{-8}$ Torr. The Meissner trap is kept filled during sputtering. After pumping down, high purity Ar, additionally purified with an oxygen trap, is admitted into the chamber to a pressure of ~ 2.5 mTorr. Up to eight substrates are mounted on a sample positioning plate that can be rotated back and forth over chosen targets under computer control. Before each sample is sputtered, the rate of sputtering of each target is checked with a quartz crystal thickness monitor, which can be positioned over each of the chosen targets. The sputtering sequence, and the times the substrate is held over each target, are computer controlled.

Each sample's resistance is measured using a potentiometer bridge circuit with a superconducting quantum interference device (SQUID) null detector [25]. To measure typical sample resistances ~ $10^{-8}$ $\Omega$, the bridge circuit has a sensitivity ~ $10^{-(11,12)}$ $\Omega$. The value of the reference resistor is known to 1%.

After the sample resistances have been measured, a Dektak surface profiler is used to measure the widths of the two ~ 1.1 mm wide Nb cross-strips. The area A ~ 1.2 mm$^2$ is then the product of the two widths. Comparing repeated measurements of the same strips by a given student, and independent measurements by other students, gives a typical uncertainty in A ~ 5%. The uncertainty in A is the largest uncertainty in each individual total specific resistance, AR$_T$. As illustrated in the figures below, however, variations in the values of AR$_T$ for nominally identical samples can be as large as 20%. We attribute these variations to real differences in the samples, rather than to either measurements of A or of R. The precise sources of these differences are not clear—presumably they involve a combination of differences in bulk and interface parameters throughout the sputtered multilayers.

3. Sample Structures for determining 2AR$_{M1/M2}$.

3.1. Two non-magnetic metals, N1/N2. All published direct measurements of 2AR$_{N1/N2}$ for non-magnetic pairs N1, N2 were made at 4.2K and used the crossed superconductor technique. In addition to 2AR$_{N1/N2}$, the analysis begins with two other unknowns, the resistivities $\rho_1$ and $\rho_2$ of N1 and N2. However, these can be determined by independently measuring them on thin (typically $\geq$ 200 nm thick) films of N1 and N2 sputtered in the same way as the multilayers used to determine 2AR$_{N1/N2}$. Such measurements leave only 2AR$_{N1/N2}$ as a true unknown. Two methods have been used to find 2AR$_{N1/N2}$.

3.1.1. Method #1, described in ref. [12], is the most frequently used. The basic idea is simple, the multilayer is divided into $n$ pairs of equal thickness layers $t_{N1} = t_{N2} = t_T/2n$, and total thickness of the multilayer is fixed at $t_T = $ 360, 540, or 720 nm. The sample has the form Nb(100)/Cu(10)/Co(10)/[N1($t_{N1}$)/N2($t_{N2}$)]$_n$/Co(10)/Cu(10)/Nb(100), where all thicknesses are in nanometers. From past experience, the two Cu layers become superconducting by the proximity effect with the adjacent superconducing Nb, but the two 10 nm thick Co layers eliminate any proximity effect on the [N1($t_{N1}$)/N2($t_{N2}$)]$_n$ multilayer of primary interest. The two Co layers are also far enough apart to eliminate any significant magnetoresistance (MR). As $n$ is increased, the contribution to the total specific resistance, AR$_T$, from scattering in the bulk of each metal remains constant, since the total thickness of each metal stays at just half of the $t_T$. If we fix $t_T$, and associate with the interfaces all of the changes in AR$_T$ with $n$, we obtain:

$$AR_T = 2AR_{Nb/Co} + 2\rho_{Co}(10) + AR_{Co/N1} + AR_{Co/N2} + \rho_{N1}(t_T/2) + \rho_{N2}(t_T/2) - AR_{N1/N2} + n(2AR_{N1/N2}). \quad (1)$$

Eq. (1) predicts that the data should grow linearly with $n$, the slope of the straight line should be just 2AR$_{N1/N2}$, and the ordinate intercept should be the sum of the first seven terms. The last of these terms, AR$_{N1/N2}$, arises because the actual sample has only (2$n$-1) N1/N2 interfaces. In earlier papers we often neglected it, since it affects only the intercept and is rarely large enough to be important. To check the internal consistency of the analysis, all of the first six terms are usually measured independently (except in a case where the sum AR$_{Co/N1}$ + AR$_{Co/N2}$ was approximated [24]) and the resulting sum is compared to the predicted intercept. So far, the two values have been consistent to within mutual uncertainties.

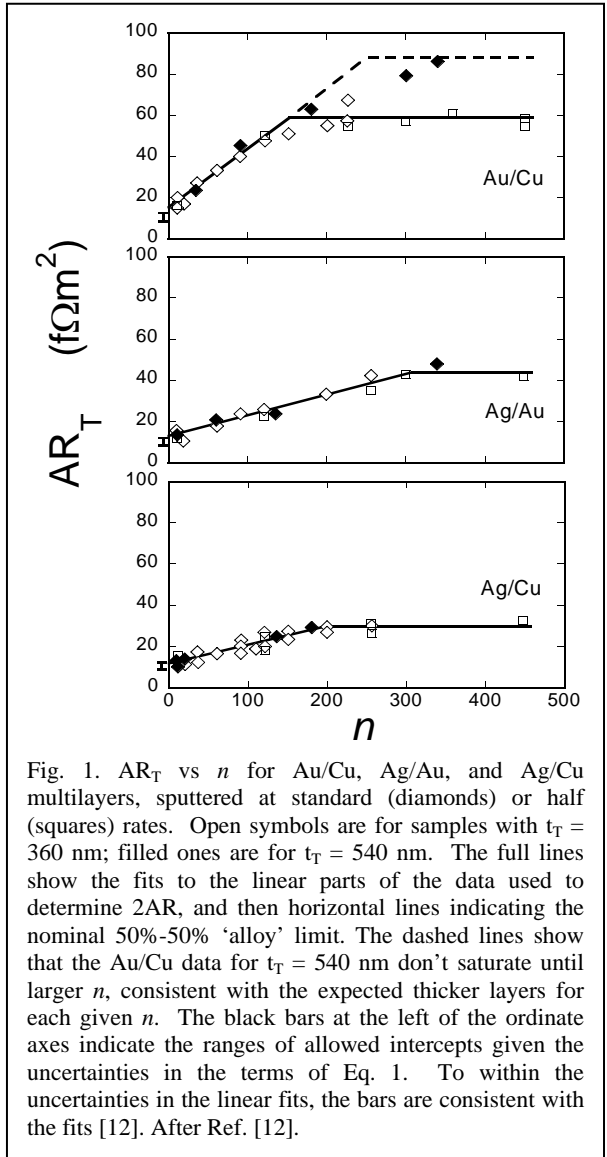

Fig. 1. $AR_T$ vs $n$ for Au/Cu, Ag/Au, and Ag/Cu multilayers, sputtered at standard (diamonds) or half (squares) rates. Open symbols are for samples with $t_T$ = 360 nm; filled ones are for $t_T$ = 540 nm. The full lines show the fits to the linear parts of the data used to determine 2AR, and then horizontal lines indicating the nominal 50%-50% 'alloy' limit. The dashed lines show that the Au/Cu data for $t_T$ = 540 nm don't saturate until larger $n$, consistent with the expected thicker layers for each given $n$. The black bars at the left of the ordinate axes indicate the ranges of allowed intercepts given the uncertainties in the terms of Eq. 1. To within the uncertainties in the linear fits, the bars are consistent with the fits [12]. After Ref. [12].

Examples of data sets for the three pairs of noble metals are shown in Fig. 1. As expected from Eq. 1, the values of $AR_T$ initially grow linearly with $n$. But, unlike the prediction of Eq. 1, the data for large $n$ deviate from this linear growth, bending over and eventually saturating. We attribute this saturation to overlap of finite thickness interfaces, a phenomenon neglected in Eq. 1. In real samples, the interfaces are never perfectly flat and the metals are never perfectly separated at such a flat interface. Rather, there is a finite thickness of intermixing, $t_I$. When the intended layer thickness, $t_{N1} = t_{N2} = t_T/2n$ becomes comparable to, or smaller than, $t_I$, the sample should first approach, and eventually become, a random 50%-50% alloy of N1 and N2. In the simplest case, where the alloy resistivity times $t_T$ is larger than $AR_T$ for values of $n$ in the linear region, the data smoothly transition to this alloy value as shown in Fig. 1.

3.1.2. Method #2, described in ref. [13], involves constructing an exchange-biased spin-valve (EBSV),

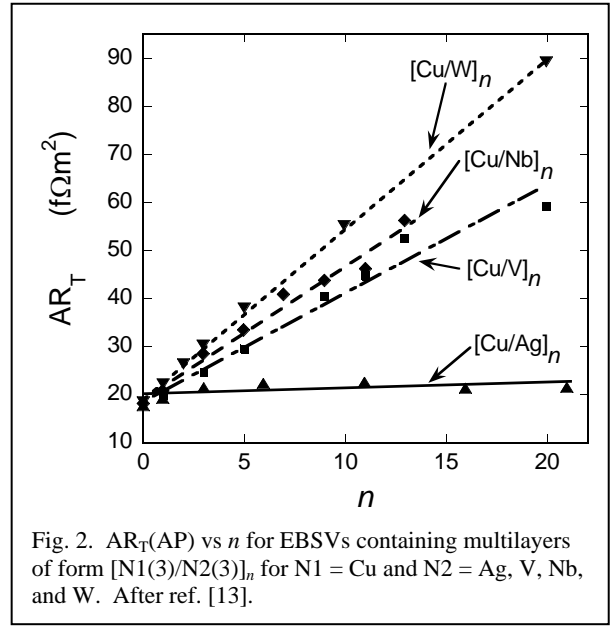

Fig. 2. $AR_T$(AP) vs $n$ for EBSVs containing multilayers of form $[N1(3)/N2(3)]_n$ for N1 = Cu and N2 = Ag, V, Nb, and W. After ref. [13].

containing a multilayer with $n$ layers of fixed (usually equal) thicknesses $t_{N1} = t_{N2} = 3$ nm. The sample form is Nb(100)/Cu(10)/FeMn(8)/Py(24)/Cu(10)/$[N1(3)/N2(3)]_n$/Cu(10)/Py(24)/Cu(10)/Nb(100), again with layer thicknesses in nanometers. The antiferromagnet FeMn pins the magnetization of the adjacent Py layer to a much higher magnetic field than that needed to reverse the other 'free' Py layer. This process allows easy magnetic field control of the two important magnetic states of the Py magnetic moments--moments parallel to each other (P), or moments antiparallel (AP). In this case, a plot of $AR_T$(AP) vs $n$ should give a linear increase, with slope equal to the sum $\rho_{N1}(3) + \rho_{N2}(3) + (2AR_{N1/N2})$. Both because determination of $2AR_{N1/N2}$ from this slope now requires subtraction of the contributions from the 'bulk' of N1 and N2, and because the contributions from the high resistance Py and FeMn components lead to uncontrolled fluctuations in $AR_T$(AP) from sample to sample, the uncertainties in $2AR_{N1/N2}$ from method #2 are generally larger than those from method #1. This is especially a problem for small values of $2AR_{N1/N2}$. Fig. 2 shows examples of $AR_T$(AP) data taken using method #2. The small slope of the data for Ag/Cu is consistent with the more reliable slope for Ag/Cu in Fig. 1. The two cases (Ag/Cu[12,13] and Pd/Pt[18]) where data have been taken for a given pair with both method #1 and method #2 involved relatively small values of $2AR_{N1/N2}$ (see Table I). In both cases, the two sets of results agreed to within mutual uncertainties. The major advantage of method #2 is that it also allows determination of the probability of spin-flipping at the N1/N2 interfaces [13], a topic covered in ref. [26]

3.2. One magnetic metal, F/N. Determining the parameters for an F/N pair is more complex, since one must distinguish between parameters for transport electrons with moments along (↑) or opposite to (↓) the moment of the F-metal through which the electrons are passing. In the bulk F, the two parameters are resistivities $\rho_F\downarrow$ and $\rho_F\uparrow$, and at the interfaces $AR_{F/N}\downarrow$ and $AR_{F/N}\uparrow$. The more directly useful alternatives for CPP-MR analysis are the following [27,28]. In bulk, the enhanced resistivity $\rho_F^* = (\rho_F\downarrow + \rho_F\uparrow)/4 = \rho_F/(1-\beta^2)$ and the dimensionless bulk asymmetry parameter $\beta = (\rho_F\downarrow - \rho_F\uparrow)/(\rho_F\downarrow + \rho_F\uparrow)$. For the interfaces, the interface specific resistance: $2AR_{F/N}^* = (AR_{F/N}\downarrow + AR_{F/N}\uparrow)/2$ and the interface asymmetry parameter $\gamma = (AR_{F/N}\downarrow - AR_{F/N}\uparrow)/(AR_{F/N}\downarrow + AR_{F/N}\uparrow)$. Measuring $\rho_F$ on a separately sputtered film allows these four parameters to be reduced to three. Independent measurements of $\rho_N$ leave these three as the only unknown parameters. Initial studies of Co/Cu [7], Co/Ag [8], and Py/Cu [10,11], required combinations of F/N multilayers sufficient to determine all three parameters. But once $\beta_F$ has been found for a given F-metal sputtered in the same way as the multilayers of present interest, the parameters to be found for an F/N pair are reduced to just $2AR_{F/N}^*$ and $\gamma_{F/N}$. Although both must be determined together, in this review we focus only upon $2AR_{F/N}^*$. Information about the related values of $\gamma_{F/N}$ can be found in the appropriate references.

To determine both $2AR_{F/N}^*$ and $\gamma_{F/N}$, one must be able to measure both $AR_T(P)$ and $AR_T(AP)$. Taking $\rho_F$, $\beta_F$, and $\rho_N$ as separately measured, in principle $2AR_{F/N}^*$ and $\gamma_{F/N}$ could be determined from measurements of AR(AP) and AR(P) on either simple $[F/N]_n$ multilayers, or exchange-biased spin-valves of the form AFM/F/N/F, where AFM indicates an antiferromagnet. In practice, with simple $[F/N]_n$ multilayers one cannot always guarantee the ability to achieve $AR_T(AP)$. With EBSVs, variations in $AR_T$, arising from fluctuations in the contributions from the high resistivities of the AFM and the two F-layers, make it difficult to derive $2AR_{F/N}^*$. Thus, the most recent studies use the simple multilayers to estimate $2AR_{F/N}^*$, and then EBSV measurements of $A\Delta R = AR(AP) - AR(P)$ to estimate $\gamma_{F/N}$. The two parameters can then be combined to check that the ensuing prediction of $A\Delta R$ for the multilayer is consistent with the actual multilayer data [20-22].

4. Calculations.

In this section we give a brief overview of calculations of 2AR [24,29-34]. Each calculation requires two steps: (A) Calculate the electronic structures of the pair N1,N2 or F,N; and (B) Given these calculated electronic structures, calculate 2AR for the assumed current direction (i.e., normal to the closest-packed plane—(111) for fcc and (110) for bcc, as noted in section 1) and a chosen interface form. The interface forms we use are: (1) a 'perfect; interface—perfectly flat and with no intermixing of the two metals; and (2) a 50%-50% random alloy, two monolayers (ML) thick.

(A) Each electronic structure is calculated using the local density approximation, assuming the equilibrium crystal structure of the metal and a chosen value for the lattice parameter a. Initial studies used linear muffin-tin orbitals (LMTO) and only *spd* angular momentum states [31,33]. The most recent calculations drop the linearization, using MTO orbitals, and expand the angular momentum states to *spdf* [17,32,34]. If, as is true for several of the pairs that we describe, the crystal structures are the same (either face-centered-cubic (fcc) or body-centered cubic (bcc)), and the lattice parameters are almost the same (e.g. $\Delta a/a \leq 1\%$), then a single crystal lattice and lattice parameter can be assumed for both metals—e.g., using the average value of a. In this case, there are no adjustable parameters. If the lattice structures are the same, but the lattice parameters differ, then there is flexibility in choice of lattice parameters. At the extremes, one can assume the average lattice parameter for both metals, or else take the equilibrium lattice parameter for each metal. If the lattice structures themselves differ (e.g., fcc vs bcc), then one might assume that each metal has its equilibrium structure. Once the electronic structures have been calculated for the chosen crystal structures and lattice parameters, one is ready for the transport calculation of 2AR.

(B) To calculate 2AR, one must first choose a form for the interface. A plausible first choice is a perfect interface, i.e., one that is perfectly flat and with no intermixing between M1 and M2. Early calculations quickly showed [29] that one can only achieve rough agreement with experiment for such an interface by assuming that the scattering within the bulk M1 and M2 is diffuse. Said another way, with a perfect interface, ballistic transport within M1 and M2 leads to quantum interference and disagreement with experiment. With assumed diffuse scattering in the bulk metals, 2AR for a perfect interface can be calculated using an appropriately modified Landauer formula [29]. This same calculation can also be done with intermixed interfaces. The simplest next choice is a 50%-50% (50-50) alloy of M1 and M2 with a thickness of 2 ML (monolayers). We will see below that the calculated values of 2AR for perfect and 50-50 2ML alloyed interfaces agree surprisingly well with all published experimental data for lattice matched pairs, but that the agreements are not so good for pairs with substantially different bulk lattice parameters.

5. Values of $2AR_{M1/M2}$, and Comparisons with Calculations.

Having described how both measured and calculated values of 2AR are obtained, and shown examples of data in Figs. 1 and 2, we turn now to the measured and calculated values of 2AR.

Table 1 contains measured and calculated quantities for several lattice matched N1/N2 and F/N pairs. The uncertainties in the calculations are due to uncertainties in the Fermi energy [35]. Note that the measured values are all fairly close to the values calculated with no-free-parameters, generally falling between the calculations for perfect and 50-50 interfaces. These agreements indicate that the calculations are reliable and that we are close to fully understanding the physics of 2AR for lattice matched pairs. Ref. [24] explains why the values of 2AR in Table 1 are mostly similar for perfect and 50-50 interfaces.

Table 2 contains measured values of 2AR for both lattice mismatched pairs and pairs with different crystal structures. In this table, the few calculations are larger than the measurements, by from 50% to as much as a factor of four. Presumably, the problem is failure to correctly account for either interfacial relaxation and/or interfacial strains. Indeed, calculations of the residual resistivities of alloys using similar techniques show a sensitivity to local structure [36].

6. Summary and Conclusions.

We have collected together measured values of $2AR_{M1/M2}$ at 4.2K for a wide variety of non-magnetic/non-magnetic (N1/N2) and ferromagnetic/non-magnetic (F/N) metal pairs. In several cases, we have been able to compare those values with calculated ones. For lattice matched pairs, the agreements with calculations having no free parameters are surprisingly good—the experimental values lie near, often between, those calculated for perfect interfaces and those calculated for interfaces composed of 2 ML of a random 50%-50% alloy. For non-lattice matched pairs, the experiments and calculations agree less well, the calculations are too large by from ~ 50% to a factor of four.

7. Acknowledgments: This work was supported in part by NSF grant DMR 08-04126.

Table I: Lattice Matched Pairs. Units for 2AR, 2AR* are f$\Omega$m$^2$. Asterisks (*) by F/N pairs indicate that the quantity listed is 2AR$_{F/N}$*.

| Metals | $\Delta a/a$(%) | 2AR(exp) | Ref. | 2AR(perf) | 2AR(50-50) | Ref. |
|---|---|---|---|---|---|---|
| Ag/Au(fcc)(111) | 0.2 | 0.10 ± 0.01 | [12] | 0.09 | 0.13 | [24] |
| Co/Cu*(fcc)(111) | 1.8 | 1.02 ± 0.1 | [7] | 0.9 | 1.1 | [24] |
| Fe/Cr*(bcc)(110) | 0.4 | 1.6 ± 0.15 | [14] | 1.7 | 1.5 | [24] |
| Pd/Pt(fcc)(111) | 0.8 | 0.28 ± 0.06 | [18] | $0.40^{+0.03}_{-0.08}$ | $0.42^{+0.02}_{-0.04}$ | [24] |
| Pd/Ir(fcc)(111) | 1.3 | 1.02 ± 0.06 | [24] | 1.1 ± 0.1 | 1.13 ± 0.1 | [24] |

Table II. Non-Lattice Matched Pairs. Units for 2AR, 2AR* are f$\Omega$m$^2$. Asterisks (*) by F/N pairs indicate that the quantity listed is 2AR$_{F/N}$*

| Metals | $\Delta a/a$(%) | 2AR(exp); | Ref. | 2AR(perf) | 2AR(50-50) | Ref. |
|---|---|---|---|---|---|---|
| Ag/Cu(fcc)(111) | 12 | 0.09 ± 0.01 | [12] | 0.45 | 0.7 | [17] |
| Au/Cu(fcc)(111) | 12 | 0.30 ± 0.01 | [12] | 0.45 | 0.6 | [17] |
| Pd/Cu(fcc)(111) | 7 | 0.9 ± 0.1 | [16] | 1.5 | 1.6 | [17] |
| Pd/Ag(fcc)(111) | 5 | 0.7 ± 0.15 | [17] | 1.6 | 2.0 | [17] |
| Pd/Au(fcc)(111) | 5 | 0.45 ± 0.15 | [17] | 1.7 | 1.9 | [17] |
| Pt/Cu(fcc)(111) | 8 | 1.5 ± 0.1 | [16] | | | |
| Ni/Cu*(fcc)(111) | 2.5 | 0.36 ± 0.06 | [23]. | 0.74 | | [30] |
| V/Cu(bcc/fcc) | | 2.3 ± 0.3 | [13] | | | |
| Nb/Cu(bcc/fcc) | | 2.4 ± 0.3 | [13] | | | |
| W/Cu(bcc/fcc) | | 3.1 ± 0.2 | [13] | | | |
| Ru/Cu(hcp/fcc) | 3 | ~ 2.2 | [15] | | | |
| Py/Cu*(fcc)(111) | 2.5 | 1.0 ± 0.1 | [10,11] | | | |
| Co/Ru*(hcp//hcp)(0001) | 5 | ~ 1 | [15] | | | |
| Co/Al*(fcc)(111) | 13 | 11.4 ± 0.3 | [20,22] | | | |
| Co$_{91}$Fe$_9$/Al*(fcc)(111) | 13 | 10.6 ±0.6 | [20,22] | | | |
| Py/Al*(fcc)(111) | 13 | 8.5 ± 1 | [19,22] | | | |
| Fe/Al*(bcc/fcc) | | 8.4 ± 0.6 | [20,22] | | | |
| Co/Pt*(fcc)(111) | 10 | 1.7 ± 0.3 | [21] | | | |
| Fe/V*(bcc)(011) | 5 | 1.4 ± 0.2 | [21] | | | |
| Fe/Nb*(bcc)(011) | 13 | 2.8 ± 0.4 | [21] | | | |
| Py/Pd*(fcc)(111) | 10 | 0.4 ± 0.2 | [21] | | | |